\documentclass{webofc}
\usepackage[varg]{txfonts}   

\usepackage{bm}
\usepackage{braket}

\begin{document}
\title{Estimation of compositeness with correction terms}
%
%

\author{\firstname{Tomona} \lastname{Kinugawa}\inst{1}\fnsep\thanks{\email{kinugawa-tomona@ed.tmu.ac.jp}} \and
        \firstname{Tetsuo} \lastname{Hyodo}\inst{1}\fnsep\thanks{\email{hyodo@tmu.ac.jp}} 
}

\institute{Department of Physics, Tokyo Metropolitan University, Hachioji 192-0397, Japan
          }

\abstract{
The compositeness $X$ is defined as the probability to observe the composite structure such as the hadronic molecule component in a bound state. One of the model-independent approaches to calculate $X$ is the weak-binding relation. However, when the scattering length $a_{0}$ is larger than the radius of the bound state $R$, the central value of the compositeness $X$ becomes larger than unity, which cannot be interpreted as a probability. For the systems with $a_{0}>R$, we need to estimate the compositeness with the correction terms. For the reasonable determination of the compositeness, we first present the quantitative estimation of the correction terms. Because the exact value of the compositeness should be contained in its definition domain $0\leq X\leq 1$, we propose the reasonable estimation method with the uncertainty band by excluding the region outside of the definition domain of the compositeness. We finally estimate the compositeness of physical systems, and obtain the result which we can interpret as the fraction of the composite component.
}
\maketitle
%

\section{Introduction} 
\label{intro}
Almost all hadrons are considered to be $qqq$ or $q\bar{q}$ states in the constituent quark models. However, some hadrons are expected to have an extraordinary structure, and called exotic hadrons. In recent experiments in the heavy quark sector, candidates for exotic hadrons have been observed, as represented by the $X(3872)$~\cite{Belle:2003nnu}. One possible component of the candidates for the exotic hadrons is the hadronic molecule, which is a weakly bound state of hadrons.

We can quantitatively characterize the internal structure of the state by compositeness whether it is a hadronic molecule dominant (composite dominant) state or not~\cite{Hyodo:2013nka}. The compositeness $X$ is defined as the probability to find the hadronic molecule component in the normalized wavefunction of the bound states $\ket{\Psi}$, $X=|\braket{{\rm molecule}|\Psi}|^{2}$. Here $\ket{\rm molecule}$ is the schematic notation of the hadronic molecule component. We can determine $X$ by using the weak-binding relation~\cite{Weinberg:1965zz,Kamiya:2016oao}:
\begin{align}
a_{0}&=R\left\{\frac{2X}{1+X}+\mathcal{O}\left(\frac{R_{\rm typ}}{R}\right)\right\},
\label{eq:wbr}
\end{align}
where $a_{0}$ is the scattering length and $R\equiv \sqrt{2\mu B}$ is the radius of the bound state, determined by the binding energy $B$ and the reduced mass $\mu$. Taking into account the range correction to the weak-binding relation~\cite{Kinugawa:2022fzn}, we define $R_{\rm typ}$ as the largest one among the length scale of the interaction $R_{\rm int}$ and those in the effective range expansion except for $a_{0}$:
\begin{align}
R_{\rm typ}=\max\{R_{\rm int},|r_{e}|,|P_{s}/R^{2}|,\cdots\},
\end{align}
where $r_{e}$ is the effective range and $P_{s}$ is the shape parameter (for more details, see Sec.~III in Ref.~\cite{Kinugawa:2022fzn}).

 When we consider sufficiently shallow bound states with $R\gg R_{\rm typ}$, the correction terms of the weak-binding relation $\mathcal{O}(R_{\rm typ}/R)$ are negligible, and the compositeness $X$ is determined only from the observables $a_{0}$ and $R$. Thanks to this universal feature, the weak-binding relation has been utilized as a model-independent approach to calculate $X$. However, naive application of Eq.~\eqref{eq:wbr} without the correction terms sometimes contradicts the definition domain of the compositeness, $0\leq X\leq 1$. For example, the compositeness of the deuteron $d$ is given as $X=1.68$ with $a_{0}=5.42$ fm (taken from CD-Bonn potential~\cite{Machleidt:2000ge}) and $B=2.22$ MeV (taken from PDG~\cite{ParticleDataGroup:2020ssz}). This problem is discussed in Refs.~\cite{Li:2021cue,Song:2022yvz,Albaladejo:2022sux} in connection with the effective range. To avoid this contradiction, here we propose a reasonable estimation method of the compositeness with the uncertainty which arises from the correction terms $\mathcal{O}(R_{\rm typ}/R)$ in Eq.~\eqref{eq:wbr}.


\section{Estimation of compositeness with correction terms}
\label{sec:estimation}
\subsection{Importance of correction terms}
\label{subsec:uncertainty}

Let us consider the relation of the scattering length $a_{0}$ and the radius $R$ by neglecting the correction terms  $\mathcal{O}(R_{\rm typ}/R)$. Because $X$ is the probability to find the composite component in a bound state, it is defined within $0\leq X\leq 1$. It follows from this relation that $2X/(1+X)\leq1$. Therefore, 
to satisfy the weak-binding relation~\eqref{eq:wbr} without the correction terms $a_{0}=R[2X/(1+X)]$, $R$ should be larger than $a_{0}$. However, there are some systems with $a_{0}>R$ which give $X>1$ as mentioned above. In such cases, we cannot interpret $X$ as the probability. This problem originates in the assumption of neglecting the correction terms. 
For the systems with $a_{0}>R$, it is expected that the weak-binding relation holds by taking into account the correction terms $\mathcal{O}(R_{\rm typ}/R)$,
because $2X/(1+X)+\mathcal{O}(R_{\rm typ}/R)>1$ can be realized for $0\leq X\leq 1$. Therefore, it is necessary to develop a quantitative estimation method of the correction terms to obtain $X\leq1$ for the systems with $a_{0}>R$.

\subsection{Estimation of uncertainty band}
\label{subsec:uncertainty2}
From the discussion in Sec.~\ref{subsec:uncertainty}, we propose the  estimation method of the compositeness $X$ with introducing the contribution from the correction terms $\mathcal{O}(R_{\rm typ}/R)$. As discussed in Ref.~\cite{Kamiya:2016oao}, the correction terms $\mathcal{O}(R_{\rm typ}/R)$ can be estimated quantitatively as the dimensionless quantity $\xi$:
\begin{align}
\xi=\frac{R_{\rm typ}}{R}.
\end{align}
We then determine the upper and lower boundaries of the estimated compositeness $X_{u}$ ($X_{l}$) as
\begin{align}
X_{u}(\xi)&=\frac{a_{0}/R+\xi}{2-a_{0}/R-\xi},\\
X_{l}(\xi)&=\frac{a_{0}/R-\xi}{2-a_{0}/R+\xi},
\end{align}
for $0\leq \xi\leq 1$. It is expected that the exact value of $X$ is contained within $X_{l}\leq X\leq X_{u}$.

\begin{figure}[h]
\centering
\includegraphics[width=10cm,clip]{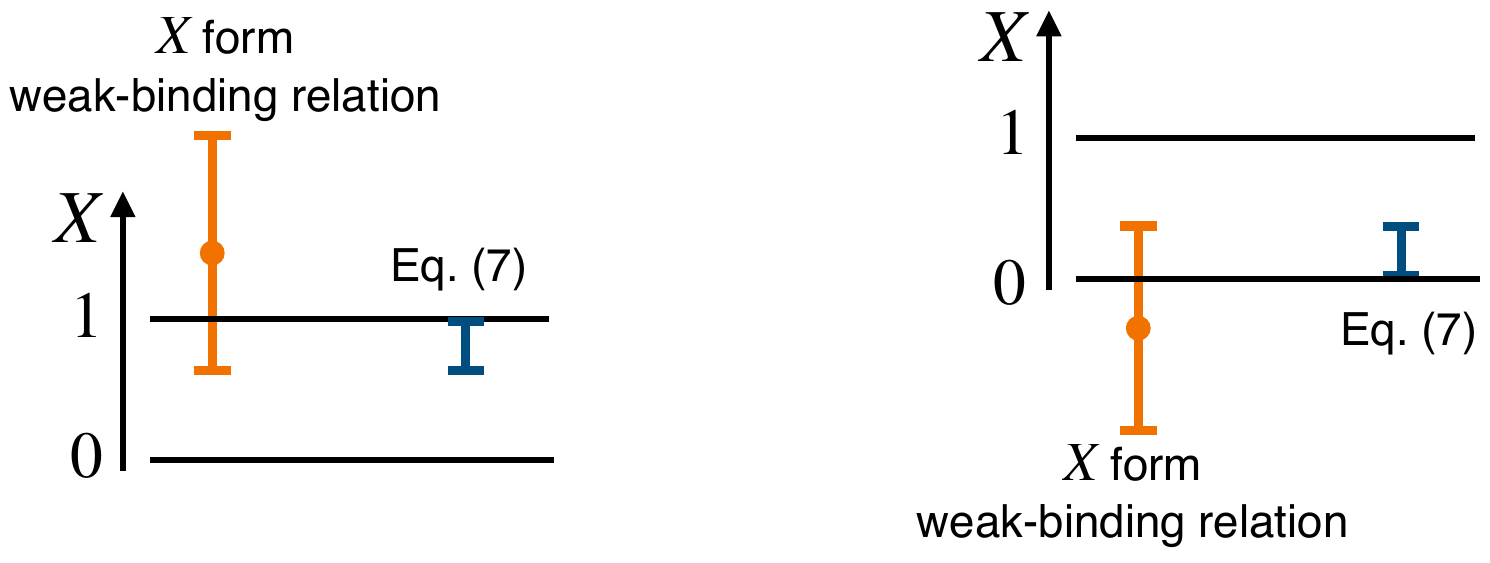}
\caption{Schematic illustration of the definition of the uncertainty band~\eqref{eq:ebar}. The left panel shows the case for $X_{u}>1$ ($\bar{X}_{u}=1$), and the right shows that for $X_{l}<0$ ($\bar{X}_{l}=0$).}
\label{fig:uncertainty}
\end{figure}

Numerically, 
$X_{u}$ and $X_{l}$ can go beyond the definition domain of the compositeness $0\leq X\leq 1$, depending on the values of $a_{0},R$ and $\xi$. However, the results $X\geq 1$ and $X\leq 0$ do not make sense, because the exact value of $X$ is not contained there. Therefore, we define 
\begin{align}
\bar{X}_{u}=\min\{X_{u},1\},\quad \bar{X}_{l}=\max\{X_{l},0\} ,
\end{align}
to restrict the uncertainty band of the compositeness within the definition domain of $X$:
\begin{align}
\bar{X}_{l}\leq X\leq \bar{X}_{u},
\label{eq:ebar}
\end{align}
as illustrated in Fig.~\ref{fig:uncertainty}. We regard this uncertainty band~\eqref{eq:ebar} as the estimated compositeness and discuss the internal structure of the bound state with it. It is clear that the estimated compositeness with the uncertainty band~\eqref{eq:ebar} is restricted within $0\leq X\leq 1$, and we can interpret $X$ as the probability. More details about the estimation of $X$ are discussed in Sec.~III and IV in Ref.~\cite{Kinugawa:2022fzn}.

\section{Application to physical systems} 
\label{sec:apply}
Now we estimate the compositeness $X$ of the actual physical systems with the uncertainty estimation discussed in Sec.~\ref{subsec:uncertainty2}. We consider the deuteron, $X(3872)$, $D^{*}_{s0}(2317)$, $D_{s1}(2460)$, $N\Omega$ dibaryon, $\Omega\Omega$ dibaryon, ${}^{3}_{\Lambda}{\rm H}$, and ${}^{4}{\rm He}$ dimer. The deuteron $d$ in the $p$-$n$ scattering is chosen as the typical observed hadron. $X(3872)$ in the $D^{0}$-$\bar{D}^{*0}$ scattering, $D^{*}_{s0}(2317)$ in the $D$-$K$ scattering, and $D_{s1}(2460)$ in the $D^{*}$-$K$ scattering are the candidates for the exotic hadrons which are experimentally observed~\cite{ParticleDataGroup:2020ssz}. $N\Omega$ and $\Omega \Omega$ dibaryons are the states obtained by the lattice QCD calculation~\cite{HALQCD:2018qyu,Gongyo:2017fjb}. We can apply the weak-binding relation not only to the hadron systems but also to the nuclei and atomic systems. ${}^{3}_{\Lambda}{\rm H}$ in the $\Lambda$-$d$ scattering is an example of nuclei, and ${}^{4}{\rm He}$ dimer which is the weakly bound state of ${}^{4}{\rm He}$ atoms is an example in the atomic systems. 

 \begin{table}
 \caption{The physical quantities and the compositeness $X$ with the uncertainty band~\eqref{eq:ebar}. u, mK and  B.R. stand for the atomic mass unit, millikelvin and the Bohr radius. \label{tab:mass}}
  \begin{tabular}{ccccccc} \hline
    bound state & $B$ & $a_0$ & $r_e$ & $R_{\rm int}$ & Compositeness $X$ \\ \hline \hline
    $d$ & $2.22$ MeV & 5.42 fm & 1.75 fm & $1.43$\ fm & $0.74\leq X\leq 1$\\ 
      $X(3872)$ & 0.018 MeV &28.5 fm & $-5.34$ fm & $1.43$\ fm & $0.53\leq X\leq 1$\\ 
       $D^{*}_{s0}(2317)$ & 44.8 MeV & 1.3 fm & $-$0.1 fm & 0.359 fm & $0.81\leq X\leq 1$\\ 
        $D_{s1}(2460)$ & 45.1 MeV & 1.1 fm & $-$0.2 fm & 0.359 fm & $0.55\leq X\leq 1$\\ 
        $N\Omega$ dibaryon  & $1.54$\ MeV & 5.30\ fm&1.26\ fm & $0.676$ fm & $0.80\leq X\leq 1$\\ 
        $\Omega \Omega$ dibaryon & $1.6$\ MeV  & 4.6\ fm &1.27\ fm& $0.949$ fm & $0.79\leq X\leq 1$\\ 
        ${}^{3}_{\Lambda}{\rm H}$ & $0.13$\ MeV & 16.8\ fm&2.3\ fm& $4.32$ fm & $0.74\leq X\leq 1$ \\ 
        ${}^{4}{\rm He}$ dimer & $1.30$\ mK & 189\ B.R.&13.8\ B.R.& $10.2$ B.R. & $0.93\leq X\leq 1$\\ \hline
  \end{tabular}
 \end{table} 

For the estimation of $X$ from the weak-binding relation, we need the scattering length $a_{0}$, the reduced mass $\mu$, the binding energy $B$, the effective range $r_{e}$, and the interaction range $R_{\rm int}$. The radius of the bound state is calculated by $R=\sqrt{2\mu B}$. We tabulate relevant quantities in Tab.~\ref{tab:mass}. We note that $R_{\rm int}$ is not an observable, and therefore it is determined from the theoretical consideration. The procedure to determine these physical quantities is explained in Ref.~\cite{Kinugawa:2022fzn}.

The results of the estimated compositeness with the uncertainty band~\eqref{eq:ebar} are shown in the right column in Tab.~\ref{tab:mass}. It is found that the range correction is important for the application to the $X(3872)$ and the $N\Omega$ dibaryon~\cite{Kinugawa:2022fzn}. We find that those bound states are dominated by the composite component because the lower boundaries $\bar{X}_{l}$ are larger than 0.5. 


\section{Summary}
\label{sec:summary}
The compositeness $X$ characterizes the internal structure of shallow bound states, especially for the candidates for exotic hadrons. The weak-binding relation is one of the approaches to estimate $X$. When we neglect the correction terms $\mathcal{O}(R_{\rm typ}/R)$, the weak-binding relation becomes completely model-independent. However, if the scattering length $a_{0}$ is larger than the radius of the bound state $R$, the compositeness is overestimated as $X\geq 1$ without the correction terms. To avoid this problem, we discuss the method to evaluate the correction terms $\mathcal{O}(R_{\rm typ}/R)$. We propose the estimation method of $X$ with the uncertainty band, which includes the contribution of the correction terms $\mathcal{O}(R_{\rm typ}/R)$. Our uncertainty estimation provides the compositeness in $0\leq X\leq 1$ which can be interpreted as a probability. We finally perform reasonable estimations of $X$ as shown in Tab.~\ref{tab:mass}, and find that all states which we consider are composite dominant ($\bar{X}_{l}\geq 0.5$).


%
%

%


\end{document}